# The Tropospheric Scintillation Prediction of Earth-to-Satellite Link for Bangladeshi Climatic Condition

Muhammad Sakir Hossain[1], Muhammad Abdus Samad[1]

**Abstract:** The performance of earth-to-satellite link largely depends on various environmental factors like rain, fog, cloud, and atmospheric effects like ionospheric and tropospheric scintillation. In this paper, the tropospheric scintillation of Bangladesh, a subtropical country, is predicted based on measured climatic parameters, like relative humidity, temperature. In this prediction, ITU scintillation model are used. Four major cities, named Dhaka, Chittagong, Rajshahi and Sylhet, of Bangladesh are selected for prediction of scintillation. From the simulation result, Rajshahi is found to be the most badly affected region by the scintillation fade depth (SFD), which is followed by Chittagong and the SFD is minimum in Dhaka and Sylhet. The difference in SFDs among the considered cities does not vary heavily. It is also found that the SFD varies from 3 dB to 13 dB depending on the frequency in used. Moreover, higher scintillation is found in rainy season of Bangladesh. During this period, the scintillation becomes double of the average figure.

## 1 Introduction

Satellite communication is drawing more attention nowadays than ever before due to its rapid deployments, larger bandwidths, and the capability of providing ubiquitous communications. One of the pre-requisite task of deploying satellite is link budget calculation, which involves determining elevation angle, losses like pointing loss, free space loss, fade margin for rain attenuation, cross-polarization due to rain, atmospheric scintillations, and so on. Fade margin for atmospheric disturbances are location and time dependent; for example, the attenuation due to rain depends on the amount of rain and rain rate is not same across the world. The same is true for other atmospheric problems. Rain attenuation experienced by Terrestrial links in Bangladeshi climatic condition is investigated in [1]. The effect of rain on satellite link performance in Bangladeshi sub-tropical climate has been investigated in [2 – 3]. But the effect of scintillation on earth-to-satellite link has not been investigated Bangladeshi climatic condition. The similar work has been done for Nigerian tropical climate [4], Thailand [5], and Malaysia [11 – 12].

---

[1]International Islamic University Chittagong, Bangladesh;
 Emails: shakir.rajbd@yahoo.com; asamadece@gmail.com

Scintillation is a sudden change of amplitude and phase of a signal. A signal experiences scintillation when it travels through a region of rapid spatial and temporal variations of dielectric parameters. Scintillation of a satellite signal does occur in ionosphere and troposphere. The signal with frequency below 3GHz is affected by the ionospheric scintillation, but the signal above 3GHz frequency experiences scintillation caused by troposphere. Since almost all of the frequency bands used in satellite communication are higher than 3 GHz, the tropospheric scintillation is a dominant effect on earth-to-satellite link. Tropospheric scintillation is the result of sudden change in refractive index caused due to the variation of temperature, water vapor content, and barometric pressure.

The prediction of scintillation is very important in satellite link budget calculation, particularly at frequencies above 10 GHz. In this paper, the tropospheric scintillation is predicted using ITU scintillation prediction model. The predicted scintillation intensity is analyzed with respect to different parameters like elevation angle, frequency, antenna diameter and so on.

## 2  Scintillation Prediction Model

Scintillation intensity prediction is usually performed using meteorological parameters measured at the ground for estimating the reference intensity. The change in refractive index in the thin layer of troposphere, which experiences the actual turbulence causing scintillation, is estimated using the ground meteorological information. There are a number of models for predicting scintillation fade like Karasawa Prediction Model [6], ITU Model [7], OTUNG Model [8], and Ortgies Models [9]. Of these, ITU model is the most widely used across the world for its accuracy in prediction. In this paper, ITU model is used for estimation of scintillation fade of different regions of Bangladesh.

ITU model estimates amplitude scintillation differently depending on the earth station elevation angle. The procedure for estimating scintillation of antenna elevation angle greater than 4º is different from that of elevation angle below 4º. Since the usual elevation angle is equal to or greater than 5º, the scintillation prediction of Bangladeshi climate will be done here considering elevation angle of greater than 4º. This model is based on monthly or longer averages of temperature $t$ (ºC) and relative humidity $H$, and reflects the specific climate conditions of the site. As the averages of $t$ and $H$ vary with season, distributions of scintillation fade depth exhibit seasonal variations, which may also be predicted by using average of $t$ and $H$ in the method. Values of $t$ and $H$ can be obtained from weather information for the site(s) in question. It is recommended to use this model at frequencies up to at least 20 GHz.

Parameters required for the method include:
- $t$: average surface ambient temperature(°C) at the site for a period of one month or longer;
- $H$: average surface relative humidity (%d) at the site for a period of one month or longer;
- $f$: operating frequency (GHz), where 4 GHz $\leq f \leq$ 20 GHz;
- $\theta$: earth station antenna elevation angle, where $\theta \geq 4°$;
- $D$: Earth station antenna diameter;
- $\eta$: antenna efficiency.

The steps of estimating scintillation fade are given below:

**Step 1:** Compute the saturation water vapour pressure, $e_s$ [hPa],

$$e_s = 6.11 e^{\frac{19.7t}{t+273}}. \tag{1}$$

**Step 2:** Compute the wet term of the radio refractivity, $N_{wet}$

$$N_{wet} = \frac{3730 \times H \times e_s}{(t+273)^2} \quad [\text{ppm}]. \tag{2}$$

**Step 3:** The standard deviation of the signal amplitude, $\sigma_{ref}$ which is used as reference is calculated as follows:

$$\sigma_{ref} = 3.6 \times 10^{-3} + 10^{-4} \times N_{wet} \quad [\text{dB}]. \tag{3}$$

**Step 4:** The effective path length $L$ is

$$L = \frac{2h_L}{\sqrt{\sin^2 \theta + 2.35 \times 10^{-4}} + \sin \theta} \quad [\text{m}]. \tag{4}$$

where $h_L$ is the height of the turbulent layer; the value to be used is $h_L$ = 1000 m.

**Step 5:** The effective antenna diameter is calculated as follows:

$$D_{eff} = \sqrt{n}D \quad [\text{m}]. \tag{5}$$

**Step 6:** Compute the averaging factor of antenna as follows:

$$g(x) = \sqrt{3.86(x^2+1)^{11/12} \sin\left[\frac{11}{6}\arctan\frac{1}{x}\right] - 7.08x^{5/6}}. \tag{6}$$

with $x = 1.22 D_{eff}^2 \frac{f}{L}$, where $f$ is the carrier frequency [GHz].

**Step 7:** Compute the standard deviation of the signal for the considered period and propagation path:

$$\sigma = \sigma_{eff} f^{7/12} \frac{g(x)}{(\sin\theta)^{1.2}} . \qquad (7)$$

**Step 8:** Calculate the time percentage factor, $a(p)$, for the time percentage $p$, of concern in the range $0.01 < p \le 50$:

$$a(p) = -0.061(\log_{10} p)^3 + 0.072(\log_{10} p)^2 - 1.71\log_{10} p + 3.0 . \qquad (8)$$

**Step 9:** Calculate the scintillation fade depth for the time percentage $p$ by:

$$A_s(p) = a(p)\sigma \quad [\text{dB}] . \qquad (9)$$

## 3  Scintillation Fade Depth Prediction

The prediction of scintillation fade depth using ITU model requires the average temperature and relative humidity over a month or longer than that. The data used in this prediction are collected from Bangladesh Agriculture Research Council [10]. The average of relative humidity and temperature of last 40 years from 1971 to 2010 are used. The average relative humidity, the maximum temperature, and the minimum temperature of different months are given in **Tables 1**, **2** and **3**. Four major cities of four regions, named Dhaka, Chittagong, Rajshahi, and Sylhet, are selected for the prediction of tropospheric scintillation.

**Table 1**
*Average Relative Humidity of four major cities of Bangladesh.*

| SL | CITY | RELATIVE HUMIDITY (%) | | | | | | | | | | | |
|---|---|---|---|---|---|---|---|---|---|---|---|---|---|
| | | Jan | Feb | March | April | May | June | July | Aug | Sept | Oct | Nov | Dec |
| 1 | Dhaka | 70 | 64 | 62 | 71 | 77 | 83 | 84 | 83 | 83 | 79 | 73 | 72 |
| 2 | Chittagong | 73 | 70 | 72 | 77 | 80 | 84 | 86 | 85 | 84 | 82 | 78 | 75 |
| 3 | Rajshahi | 76 | 69 | 61 | 64 | 74 | 83 | 87 | 86 | 86 | 82 | 77 | 76 |
| 4 | Sylhet | 74 | 68 | 67 | 76 | 81 | 87 | 87 | 86 | 86 | 83 | 77 | 75 |

The scientillation fade depth will be predicted considering different factors like elevation angle, percentage of time, frequency in use, antenna diameter, and so on. Since the temperature of a day varies over a wide range, so the scientillation resulting from the relative humidity and temperature also changes over time. In this simulation work, the average maximum temperature and the average minimum temperature are used. Thus, the SFD will have two values for the same set of parameters, maximum SFD and minimum SFD. Throughout this paper, the earth station antenna effciency is assumed as 0.5 and the antenna diameter is 8 m, the diameter of earth station used as the hub station of star VSAT configuration. Very low elevation angle is used to find the worst case scenario because very low elevation angle causes signal to travel long atmosphere which results in more scintillation. Fig. 1 shows the dependecy of SFD over elevation angle of earth station antenna at Ku band (10.95 GHz) for 0.01% of time. While the maximum SFD, due to the maximum temperature, of 8.5 dB is found at Rajshahi, the minimum figure of 5 dB is observed at Dhaka and Sylhet. There are about 40 variation between these figures. The SFDs of different cities are very close to each others. It is clear from the figure that the SFD changes with elevation angle exponentially. SFD becomes below 2 dB at 15º elevation angle. The north-west and south-east regions of Bangladesh are Tetulia (with latitude: 26.5º and longitude: 88.34º) Tekhnaf(with latitude:20.86º and longitude:92.23º), respectively.

**Table 2**
*Average maximum temperature of four major cities of Bangladesh.*

|   | CITY | MAXIMUM TEMPARATURE (ºC) | | | | | | | | | | | |
|---|------|------|------|------|------|------|------|------|------|------|------|------|------|
|   |      | Janu | Febr | March | April | May | June | July | Aug | Sept | Oct | Nov | Dec |
| 1 | Dhaka | 25 | 28 | 32 | 34 | 33 | 32 | 32 | 32 | 32 | 32 | 29 | 26 |
| 2 | Chittagong | 26 | 28 | 31 | 32 | 32 | 32 | 31 | 31 | 32 | 32 | 30 | 27 |
| 3 | Rajshahi | 24 | 28 | 33 | 36 | 35 | 34 | 32 | 33 | 32 | 32 | 29 | 26 |
| 4 | Sylhet | 25 | 28 | 31 | 31 | 31 | 31 | 31 | 32 | 31 | 31 | 29 | 27 |

**Table 3**
*Average minimum temperature of four major cities of Bangladesh.*

|   | CITY | MINIMUM TEMPARATURE (ºC) | | | | | | | | | | | |
|---|------|------|------|------|------|------|------|------|------|------|------|------|------|
|   |      | Janu | Febr | March | April | May | June | July | Aug | Sept | Oct | Nov | Dec |
| 1 | Dhaka | 13 | 16 | 21 | 24 | 25 | 26 | 26 | 26 | 26 | 24 | 19 | 14 |
| 2 | Chittagong | 14 | 16 | 20 | 24 | 25 | 25 | 25 | 25 | 25 | 24 | 20 | 16 |
| 3 | Rajshahi | 11 | 13 | 18 | 23 | 24 | 26 | 26 | 26 | 26 | 23 | 18 | 13 |
| 4 | Sylhet | 13 | 15 | 18 | 21 | 23 | 24 | 25 | 25 | 25 | 23 | 19 | 14 |

All regions of Bangladesh fall inside these latitudes and longitudes. Thus, the SFD can be kept 2 dB if the satellite longitude is between 152º and 157º. Using satellite latitude greater than 157º, the SFD can be reduced further but this will increase slant range which will cause the rise of attenuation due rain. Thus, a tradeoff must be made between these two factors.

The SFD does not remain same throughout the year, rather; a significant variation is observed with time. This is due to the changing pattern of temperature and humidity. Fig. 2 shows the minimum SFD at three different elevation angles for different percentage of times at Ku band. At 5º elevation angle, the SFD lies between 5.2 dB to 5.7 dB, where the maximum and the minimum SFD are observed at Rajshahi, and Dhaka and Sylhet, respectively, for 0.01% of time of the year. There are 63% and 81% reduction in SFD for an increase of elevation angle by 5º and 15º, respectively. In addition, the SFD for 0.1% time is roughly 30% less compared to that at 0.01% time of year. The variation in SFD in each day is about 30%. Fig. 3 shows the maximum SFD at Ku band. Fig.3 shows the maximum SFD at Ku band.

All regions of Bangladesh fall inside these latitudes and longitudes. Thus, the SFD can be kept 2 dB if the satellite longitude is between 152º and 157º. Using satellite latitude greater than 157º, the SFD can be reduced further but this will increase slant range which will cause the rise of attenuation due rain. Thus, a tradeoff must be made between these two factors.

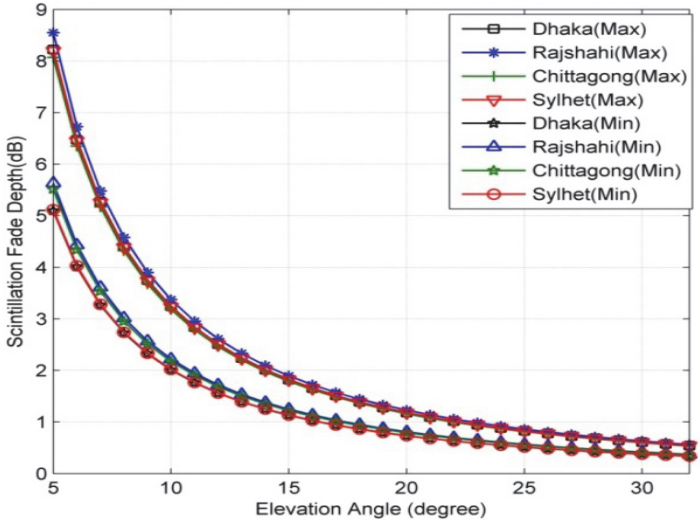

**Fig. 1** – *Annual maximum and minimum scintillation fade depth for* 0.01% *of time with frequency Ku band* (10.95 GHz).

The SFD does not remain same throughout the year, rather; a significant variation is observed with time. This is due to the changing pattern of temperature and humidity. Fig. 2 shows the minimum SFD at three different elevation angles for different percentage of times at Ku band. At 5º elevation angle, the SFD lies between 5.2 dB to 5.7 dB, where the maximum and the minimum SFD are observed at Rajshahi, and Dhaka and Sylhet, respectively, for 0.01% of time of the year. There are 63% and 81% reduction in SFD for an increase of elevation angle by 5º and 15º, respectively. In addition, the SFD for 0.1% time is roughly 30% less compared to that at 0.01% time of year. The variation in SFD in each day is about 30%. Fig. 3 shows the maximum SFD at Ku band. Fig.3 shows the maximum SFD at Ku band.

The frequency used in satellite communication plays an important role in tropospheric scintillation fade depth. As it is seen from Fig. 4 and Fig. 5, there is a positive correlation between frequency in use and the SFD. The SFD ranges from 3 dB to 13 dB for 0.01% of time. While the minimum SFD of 3 dB is found at Dhaka, the maximum figure is reported at Rajshahi which is 13 dB. The maximum SFD for 0.1% and 1% time of year are practically 9 dB and 5 dB, respectively, at Ka band. It is also found that the more difference in minimum SFD among different cities are exist compared to their maximum SFDs. For example, the minimum SFD of Chittagong at Ka band is almost equal to that of Dhaka at Ku band. However, the SFD encountered in Ka band is almost 25% and 60% higher compared to that in Ku band and C bands.

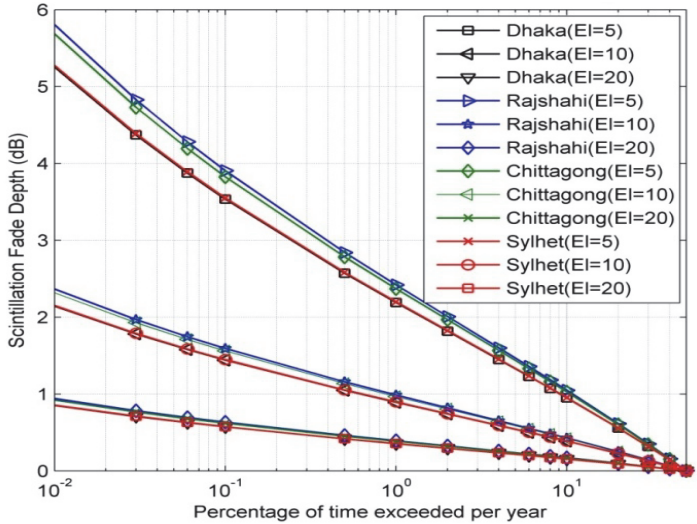

**Fig. 2** – *Annual minimum scintillation fade depth for different elevation angles with frequency Ku band.*

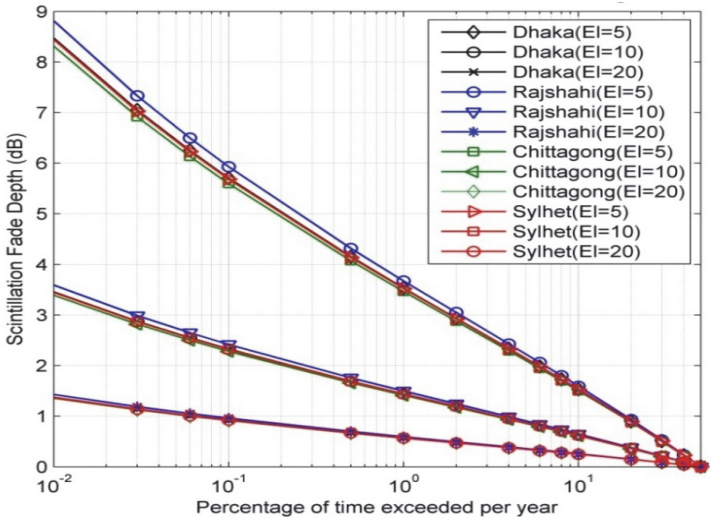

**Fig. 3** – *Annual maximum scintillation fade depth at different elevation angle in Ku band.*

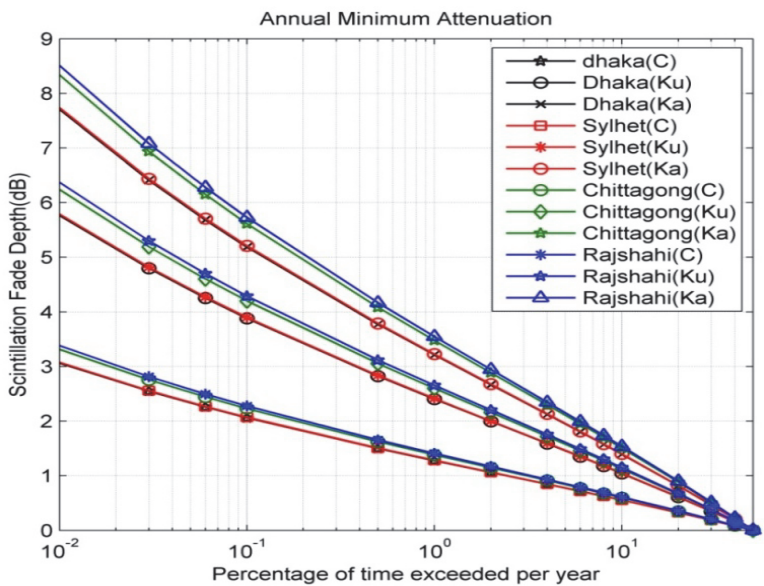

**Fig. 4** – *Annual minimum attenuation of different bands with elevation angle of* 5º.

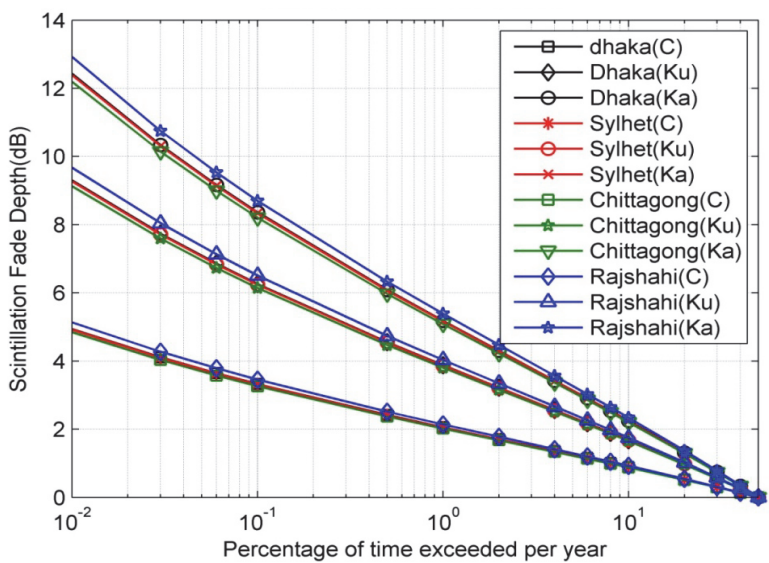

**Fig. 5** – *Annual maximum attenuation of different bands with elevation angle of* 5º.

The relative humidity and temperature varies considerably with time and space. These changes bring about variation in the corresponding SFD. Fig. 7 shows the changing pattern of average SFD throughout the twelve years in Ku and Ka bands. SFD is minimum from January to March, which is followed by a sharp rise up to the month of June until it reaches its peak value, which is around 10 dB in Ka band and 8 dB in Ku band. It maintains the same value with some slight variations between June to September, before starts declining abruptly during the rest of the months of the year. The SFD in June-September is nearly two times higher in Ka and Ku bands compared to that in January-March. During rainy sessions, the difference in SFD between Ka and Ku bands is almost 20%.

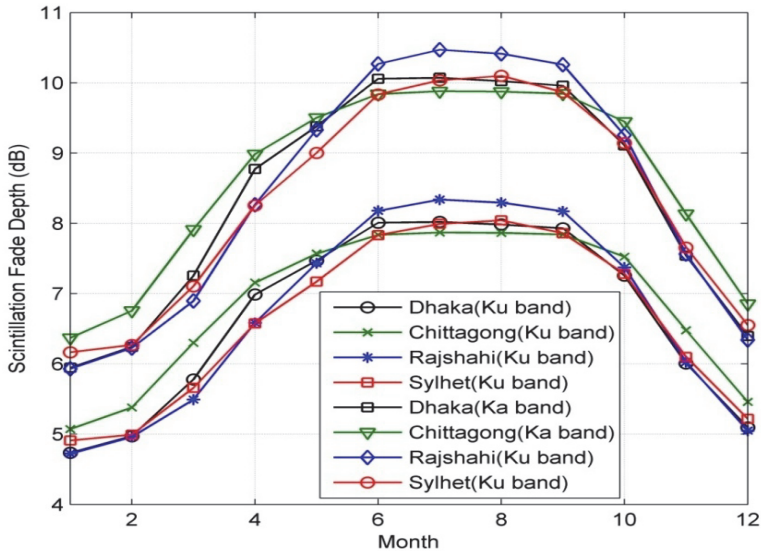

**Fig. 6** –*The variations of scintillation fade depth over the year at Ka band and elevation angle 5º.*

The effect of earth station antenna diameter on the earth-to-satellite like in Bangladeshi climatic condition is shown in Fig. 7 for Ku and Ka bands. This simulation is performed considering antenna efficiency of 0.5, 0.01% of time, and 5º elevation angle. The earth-to-satellite link using Ka band encounters more SFD compared to that of Ku band. A decrease in SFD is observed with the increase of antenna diameter. The rate of fall of SFD in Ka band is sharper than that of Ku band. The scintillation fade depth can be nullified in Ka band using an antenna having 20 m diameter, but the achieving of the same result in Ku band is possible using larger antenna whose diameter is 26 m. The interesting point is that the SFD of Ku band and Ka band can be made equal if the earth station antenna diameter is 15 m.

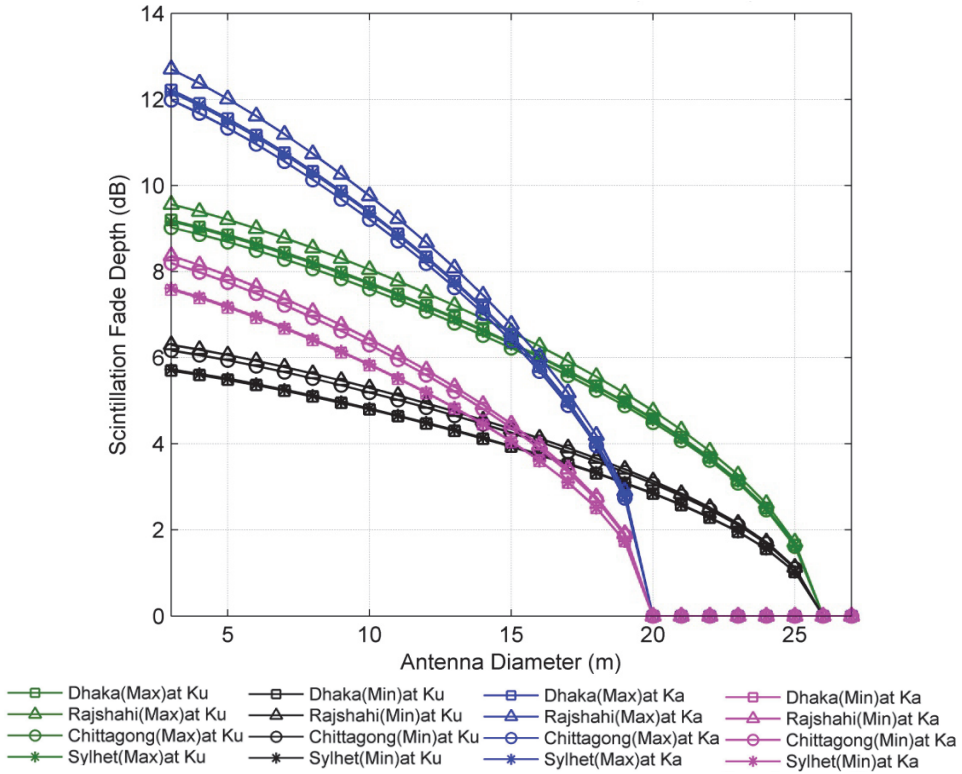

**Fig. 7** –*Variations of scintillation fade depth with antenna diameter at Ku and Ka bands.*

## 4  Conclusion

Bangladesh, as a subtropical country, is one of the rainiest regions of the world and rain directly affects the humidity and temperature, which in turn plays main role in tropospheric scintillation. From the simulation results, it is found that if the longitude of satellite is more than 152º, it is possible to use 15º elevation angle, thereby keeping the scintillation fade depth below 2 degrees at Ku band. In addition, a negative correlation is found between SFD and elevation angle. However, there is a significant change of SFD over months. The SFD from December to March is very low and less than half of the figure observed from June to September. Thus, for avoiding the huge SFD in rainy season, two different frequencies can be used. C band can be used in rainy season and Ka band in winter season. Moreover, the earth station antenna diameter of 15 meter can be used because this diameter provides the same SFD for both Ku and Ka band, thus making it easier to switch frequency.